\documentstyle[aps,prb,multicol]{revtex}
\begin{document}
\draft
\title{Self-Organized Criticality Effect on
Stability:\\ Magneto-Thermal Oscillations in a
Granular YBCO Superconductor}
\author{L.~Legrand and I.~Rosenman\\}
\address{Groupe de Physique des Solides, Unit\'e 17
Associ\'ee au CNRS,\\
Universit\'es Paris 6 et Paris 7, Tour 23, 2 place
Jussieu, 75251 Paris Cedex 5, France}
\author{R.G.~Mints}
\address{School of Physics and Astronomy,
Raymond and Beverly Sacler Faculty of Exact Sciences,\\
Tel Aviv University, Tel Aviv 69978, Israel}
\author{G.~Collin}
\address{Laboratoire L\'eon Brillouin, CEN-Saclay, 91191
Gif-sur-Yvette Cedex, France}
\date{\today}
\maketitle
\begin{abstract}
We show that the self-organized criticality of the Bean's
state in each of the grains of a granular superconductor
results in magneto-thermal oscillations preceding a series
of subsequent flux jumps. We find that the frequency of
these oscillations is proportional to the external magnetic
field sweep rate $\dot H_e$ and is inversely proportional
to the square root of the heat capacity. We demonstrate
experimentally and theoretically the universality of this
dependence that is mainly influenced by the granularity of
the superconductor.
\end{abstract}
\pacs{74.60.Ge, 74.72.Bk, 74.80.Bj}
\begin{multicols}{2}
\narrowtext
The theory of the self-organized criticality
\cite{bak1,bak2} explains the dynamics of the nonequilibrium
systems similar to sandpiles. The magnetic flux dynamics
and, in particular, the magnetization relaxation process in
a superconductor with a strong pinning potential for
vortices is an interesting and challenging field of
application for this theory \cite{tang,wang,pan}. In the
scope of the ideas of the self-organized criticality the
pinned vortices space distribution arises as a result of a
subsequent local vortex avalanches, {\it i.e.}, a series of
small local flux jumps is establishing the critical state.
The Bean critical state model \cite{bean} successfully
describes the irreversible magnetization in type-II
superconductors by introducing the critical current density
$j_c$. In the framework of the Bean model the value of the
slope of the stationary magnetic field profile is less or
equal to $\mu_0 j_c$. It makes the spatial distribution
of vortices in a superconductor with a strong pinning
potential similar to the sand particles spatial distribution
in a sandpile \cite{gen}.
\par
The stationary critical state becomes unstable under certain
conditions when the local flux jumps result in a global
flux jump driving the system to the normal state
\cite{mints1}. This instability can be preceded by a series
of magneto-thermal oscillations \cite{mints2}. These
oscillations have been reported earlier \cite{zeb,chi,lor}
but never have been studied.
\par
Each of the local vortices avalanches establishing the
critical state produces a heat pulse and a temperature rise
in the superconductor. This temperature rise decreases the
critical current density for a certain time interval and,
thus, changes the initial conditions for the subsequent
vortices avalanches. In other words, the heat pulses
produced by the local vortices avalanches result in a
correlation mechanism specific for the self-organized
criticality of magnetic flux motion in superconductors with
a strong pinning potential. In this Letter we demonstrate
that this mechanism results in the
magneto-thermal oscillations arising close to the threshold
of the superconducting state stability. We focus our study
on the dependence of the frequency of these oscillations on
the temperature and the magnetic field sweep rate in case
of a granular superconductor. We point out that the
granularity and the randomness of the properties of the
superconducting grains result in universality of this
dependence.
\par
We begin with the theoretical consideration of the critical
state stability and magneto-thermal oscillations. We
propose a one-dimensional model of a granular superconductor
treating it as a stack of superconducting slabs having the
width $2b_i\ (i=1, 2, 3, \ldots, N)$ randomly distributed
with a certain mean value $b$. We assume that there is no
electrical contact between the slabs and there is an ideal
thermal contact between them. We suppose that the external
magnetic field $H_e(t)$ is parallel to the sample surface
(${\bf H}_e$ is parallel to the $z$-axis) and the magnetic
field sweep rate $\dot H_e$ is constant. We assume that the
critical state arises simultaneously in the entire
superconductor and therefore in each of the slabs the
background magnetic, $B_i(x,t)$, and electric, $E_i(x,t)$,
fields are determined by Maxwell equations
\begin{equation}
{dB_i\over dx}=\pm\,\mu_0 j_c, \qquad
{dE_i\over dx}=\pm \dot H_e.\
\label{e1}
\end{equation}
We suppose also that close to the instability threshold most
of the slabs are saturated, {\it i.e.}, the external
magnetic field $H_e$ is higher than the Bean field
$B_p=\mu_0 j_cb$.
\par
Magneto-thermal oscillations in the critical state arise as
coupled oscillations of small perturbations of the
temperature, $\theta$, and the electric field, $\epsilon$.
The heat diffusion and Maxwell equations determine the
spatial and temporal variations of $\theta$ and $\epsilon$,
namely,
\begin{eqnarray}
C\,{\partial\theta\over\partial t}=
\kappa\,{\partial^2 \theta\over\partial x^2}+j_c\epsilon,
\label{e2}\\
\mu_0\,{\partial j\over\partial t}=
{\partial^2 \epsilon\over\partial x^2},
\label{e3}
\end{eqnarray}
where $C$ is the heat capacity, $\kappa$ is the heat
conductivity, and
\begin{equation}
{\partial j\over\partial t}={\partial j\over\partial E}\,
{\partial\epsilon\over\partial t}-
\Big\vert{\partial j_c\over\partial T}\Big\vert\,
{\partial \theta\over\partial t}.
\label{e4}
\end{equation}
\par
In the critical state the current density, $j$, is close to
$j_c$. In this region the j-E curve takes the form
\cite{gur}
\begin{equation}
j=j_c+j_1\,\ln\Bigl({E\over E_0}\Bigr),
\label{e5}
\end{equation}
where $E_0$ is the voltage criterion at which $j_c$ is
defined, $j_1$ determines the slope of the j-E curve, and
$j_1\ll j_c$.
\par
The relation given by Eq.~(\ref{e5}) was first derived in
the framework of the Anderson-Kim model
\cite{and1,kim,and2} considering the thermally activated
uncorrelated hopping of bundles of vortices $T$. It follows
from this model that $j_1\propto T$. In the framework of
the self-organized criticality the value of $j_1$ seems to
be temperature independent in the range of $T\ll T_c$
\cite{wang,pan}. We will consider the later case as it is
in a good agreement with numerous experimental data
\cite{gur}. To be more precise, we suppose that the ratio
$n=j_c/j_1\gg 1$ is temperature independent at $T\ll T_c$.
\par
Using Eqs.~(\ref{e2})~-~(\ref{e5}) we obtain the equations
to determine $\theta (x,t)\propto\exp (\gamma t)$ and
$\epsilon (x,t)\propto\exp (\gamma t)$ in the form
\begin{eqnarray}
{nE_i\over\mu_0\gamma j_c}\,\epsilon^{''}-\epsilon =
{nE_i\over j_c}\,
\Big\vert{\partial j_c\over\partial T}\Big\vert\,\theta,
\label{e6}\\
\kappa\theta^{''}+nE_i\,
\Big\vert{\partial j_c\over\partial T}\Big\vert\,\theta
-\gamma C\theta ={nE_i\over\mu_0\gamma}\,\epsilon^{''}.
\label{e7}
\end{eqnarray}
\par
Let us clarify the following calculation qualitatively.
Suppose that the initial temperature of the superconductor
$T_0$ increases by a small perturbation $\theta_0$ arising
due to a heat pulse with the energy $\delta Q_0$. This
temperature increase leads to a decrease of the
superconducting currents. The reduction of these
screening currents results in an additional flux penetration
inside the superconductor. This flux motion induces an
electric field perturbation $\epsilon_0$ producing an
additional heat release $\delta Q_1$, an additional
temperature rise $\theta_1$, and, consequently, an
additional reduction of the superconducting currents. At
certain conditions this process results in an avalanche-type
increase of the temperature and magnetic flux in the
superconductor, {\it i.e.}, in a critical state instability.
\par
The critical state is stable if the heat release,
$\delta Q$, arising in the process of electric field and
temperature perturbations development  is less than the
maximum heat flux to the coolant. The value of $\delta Q$
depends on both $\dot H_e$ and $H_e$ for the unsaturated
grains and only on $\dot H_e$ for the saturated grains. In
our experiments most of the grains are saturated in the
magnetic field region corresponding to the magneto-thermal
oscillations. Therefore, the heat release in the unsaturated
grains, $\delta Q_u$, is small compared with the heat
release in the saturated grains, $\delta Q_s$. However, the
term $\delta Q_u$ is the only magnetic field dependent term
in the heat balance equation and, thus, it determines the
value of the global flux jump field $H_j$. In other words,
the relatively small heat release in the unsaturated grains
is tuning the superconducting state in a granular
superconductor to the instability.
\par
The heat release $\delta Q$ depends on the frequency of the
magneto--thermal oscillations, $\omega ={\rm Im}\,\gamma$,
in both the saturated and unsaturated grains. Thus, to
calculate the value of $\omega$ we take into account
only the dominating heat release arising in the saturated
grains. In other words, the frequency $\omega$ is mainly
determined by the response of the active media of the
saturated grains to the small temperature and electric field
perturbations. The vortices avalanches establishing the
critical state become strongly correlated in the vicinity
of the instability threshold.
\par
Under conditions of our experiments the temperature $\theta$
is practically uniform over the cross-section of the sample.
Therefore, to solve Eq.~(\ref{e6}) we consider $\theta$ to
be constant. In a saturated slab the background electric
field $E_i=\dot H_e x$, where $x=0$ corresponds to the
middle plain. In this case Eq.~(\ref{e6}) takes the form
\begin{equation}
{n\dot H_e\over\mu_0\gamma j_c}\,x\,\epsilon^{''}-
\epsilon={n\dot H_e\theta\over j_c}\,
\Big\vert{\partial j_c\over\partial T}\Big\vert\,x
\label{e8}
\end{equation}
with the boundary conditions $\epsilon (\pm b_i)=0$. Note
that the characteristic space scale of Eq.~(\ref{e8}) is
given by
\begin{equation}
l={n\dot H_e\over\mu_0\gamma j_c}.
\label{e9}
\end{equation}
\par
The magneto-thermal oscillations exist for low values of
$\dot H_e$ where $l\ll b_i$. It allows to solve
Eq.~(\ref{e8}) using the WKB approximation that results in
\begin{equation}
\epsilon={n\dot H_e\theta\over j_c}\,
\Big\vert{\partial j_c\over\partial T}\Big\vert\,
\Biggl[x-\sqrt{b_il}\,{\sinh\bigl(2\sqrt{x/l}\bigr)
\over\cosh\Bigl(2\sqrt{x/l}\Bigr)}\Biggr].
\label{e10}
\end{equation}
\par
We integrate now Eq.~(\ref{e7}) over the cross-section of
the superconductor using Eq.~(\ref{e10}) for the electric
field $\epsilon$ and we end up with the frequency of the
magneto-thermal oscillations in the form
\begin{equation}
\omega={\dot H_e\over\sqrt{\mu_0C(T_0)T_*}},
\label{e11}
\end{equation}
where $T_0$ is the mean temperature of the oscillations
and where we determine the parameter $T_*$ as
\begin{equation}
{1\over T_*}=\sum\limits_{i=1}^N{n_i^2\over j_c^i}\,
\Big\vert{\partial j_c^i\over\partial T}\Big\vert\,b_i
\Big/\sum\limits_{i=1}^N b_i.
\label{e12}
\end{equation}
\par
Thus, the frequency of the magneto-thermal oscillations
in a granular superconductor is proportional to the
magnetic field sweep rate $\dot H_e$. The value of the
parameter $T_*\sim T_c/n^2$ is a certain constant in the
temperature range $T_0\ll T_c$. It follows then from
Eq.~(\ref{e12}) that the ratio
$\mu=\omega\sqrt{C(T_0)}/\dot H_e$ is independent on $T_0$
and $\dot H_e$, {\it i.e.}, it is a constant characterizing
the properties of the superconductor and its granular
structure.
\par
We perform an experimental study of the magneto-thermal
oscillations in a textured YBa$_2$Cu$_3$O$_{7-\delta}$
superconductor grown from the melt and heat treated after
the preparation \cite{lor2}. First, we do the measurements
using the original sample (S0) with the size of
8$\times$9$\times$5.5~mm$^3$. Next, we cut this sample in
two approximately equal parts (S1 and S2) and we measure
the magneto-thermal oscillations in S1 and S2. In this way
we check the independence of the oscillations frequency on
the size of the sample, {\it i.e.}, its intrinsic origin
influenced mainly by the granular structure of the
superconductor.
\par
We perform the magnetic characterization by DC and AC
measurements in a Quantum Design SQUID magnetometer using
the sample S3 with a size of
0.3$\times$0.5$\times$0.8 mm$^3$ (the sample S3 is cut from
the sample S2). We find the onset temperature at zero
magnetic field $T_c\approx$\ 88\ K and the AC
susceptibility transition width $\Delta T\approx$\ 5\ K.
\par
We show in Fig.~1 one quarter of the DC magnetization loops
measured for the sample S3 for several temperatures in the
magnetic field range $0<H_e<5$\ T. We find also that in
parts of the sample S0 smaller than S3 the Bean penetration
field remains the same as for S3. It means that the space
scale of the screening current loops is smaller than the
dimensions of the sub-samples, {\it i.e.}, the sample
consists of small superconducting grains. We show in the
inset in Fig.~1 the temperature variation of the critical
current extracted from the saturation value of the
magnetization. As can be seen the critical current density
decreases linearly in the range $2<T<6$\ K.
\par
The experimental setup for the magneto-thermal oscillations
measurements consists essentially of a sample holder
with a thin hollow powdered graphite column with low
thermal conductance on which the YBCO sample is maintained.
A small size (0.2 mm thick, 2 mm$^2$ surface) carbon
thermometer is glued to the sample. The entire sample
holder is maintained under a very low
pressure of He gas (2$\times$10$^{-6}$ Torr) in order to
reduce the heat leak to the coolant. The sample is first
cooled down to the starting temperature $T_s$ in zero
magnetic field. Then the field is established at a
controlled rate and the temperature variation of the sample
is measured.
\par
We measure the magneto-thermal oscillations at temperatures
in the range $2<T_0<7$\ K and the field sweep rates within
the interval $15<\dot H_e<60$\ G/s. We show in Fig.~2 a
typical sample temperature time dependence for $T_0=5$\ K
and $\dot H_e=20.4$\ G/s. At low values of the magnetic
field $H_e(t)$ there is a slow temperature increase due to
the small vortices avalanches establishing the critical
state. Above a certain magnetic field value, the sample
temperature oscillations appear with a period $\tau$ in the
range $10<\tau<70$\ s and an amplitude increasing in time.
These magneto-thermal oscillations we mainly explore and
analyze. A flux jump occurs close to the Bean field
accompanied by a temperature rise up to about 12\ K with a
characteristic time of the order of 1\ s. Then the sample
temperature relaxes to the coolant temperature with a rate
depending on the heat leak.
\par
We show in Fig.~3 the magneto-thermal oscillations frequency
dependence on the field sweep rate, for several temperatures
and different samples. As can be seen the value of $\omega$
increases linearly with $\dot H_e$ within the accuracy of
our measurements.
\par
We show in Fig.~4 the ratio
$\mu=\omega\sqrt{C(T_0)}/\dot H_e$ as a function of the
magnetic field sweep rate $\dot H_e$ for the samples S0, S1,
and S2 for different temperatures $T_0$ from the interval
$2<T_0<7$\ K. We use the relation \cite{eck,aya}
\begin{equation}
C=100\,T+5\,T^3\ ({\rm JK}^{-1}{\rm m}^{-3})
\label{e13}
\end{equation}
to calculate the heat capacity and we normalize the values
of $\mu$ by the mean value $\bar\mu$ of their distribution.
We see in Fig.~4 that the ratio $\mu$ is a constant within
the accuracy of our experiments as predicted by
Eq.~(\ref{e11}). Assuming that
$T_*^{-1}\approx n^2\,(dM/dT)/M$ we estimate the value
of $n$ as $n\approx 14.5$ which is in a good agreement with
the known experimental data \cite{gur}.
\par
To verify the uniformity of the sample temperature we
estimate the characteristic heat redistribution time
$t_h=Cd^2/\kappa$, where $d$ is the sample size. We find
that $t_h\approx 0.06$\ s using the data
$T\approx 5$\ K, $C\approx 1200$\ JK$^{-1}$m$^{-3}$,
$\kappa\approx 2$\ WK$^{-1}$m$^{-1}$, and $d\approx 1$\ cm.
This time constant is two orders of magnitude less than
both the period of the magneto-thermal oscillations and the
temperature relaxation time. Thus, the temperature of the
sample is uniform as suggested above.
\par
In conclusion, we show theoretically that the
self-organized criticality of the Bean's state in each of
the grains of a granular superconductor results in
magneto-thermal oscillations preceding flux jumping. We
study these oscillations experimentally in a granular YBCO
samples at temperatures $2<T<7$\ K and field sweep rates
$10<\dot H_e<60$\ G/s. We find both experimentally and
theoretically that the frequency of the magneto-thermal
oscillations is proportional to the magnetic field sweep
rate and is inversely proportional to the square root of
the heat capacity. We demonstrate the universality of this
dependence measured for different samples.
\par

\begin{figure}
\caption{One quarter of the magnetization loops for the
sample S3 for several temperatures. The inset shows the
temperature dependence of the critical current density.}
\label{fig1}
\end{figure}
\begin{figure}
\caption{The temperature of the sample S1 as a function of
the magnetic field $H_e$ for $\dot H_e=20.4$\ G/s and
$T_0=5.0$\ K. The inset shows the magneto-thermal
oscillations.}
\label{fig2}
\end{figure}
\begin{figure}
\caption{The dependencies $\omega (\dot H_e)$ for the
samples: S0 at $T_0=6$\ K (filled circles), S1 at
$T_0=3.8$\ K (filled triangles) and at $T_0=5.3$\ K (open
circles), and sample S2 at $T_0=5.2$\ K (open squares). The
error bars are 8\%.}
\label{fig3}
\end{figure}
\begin{figure}
\caption{The ratio $\mu=$ for the samples S0, S1 and S2 as
a function of the sweep rate for different temperatures
$T_0$ from the range $2<T_0<7$\ K. The values of $\mu$ are
normalized by the mean value of their distribution. The
error bars are 13\%.}
\label{fig4}
\end{figure}
\end{multicols}
\end{document}